\PassOptionsToPackage{pdftex}{graphicx}
\documentclass[fp,twocolumn]{jpsj3}
\usepackage{txfonts}
\usepackage{bm}
\usepackage{url}

\title{Analysis of Critical Points in a Permutation Model on Hierarchical Lattices by Real-Space Renormalization Group}

\author{Ryuki Ito$^{1}$, Taisei Matsuo$^1$, and Masayuki Ohzeki$^{1,2,3,4,*}$}
\inst{$^1$Department of Physics, Institute of Science Tokyo, Tokyo 152-8551, Japan \\
$^2$Graduate School of Information Sciences, Tohoku University, Miyagi 980-8564, Japan \\
$^3$Research and Education Institute for Semiconductors and Informatics, Kumamoto University, Chuo, Kumamoto 860-0862, Japan \\
$^4$Sigma-i Co., Ltd., Tokyo 108-0075, Japan \\
$^*$E-mail: mohzeki@tohoku.ac.jp}

\abst{The permutation model is a classical spin system where elements of the symmetric group interact with one another. The partition function of this model is directly related to the entanglement structure of random quantum circuits and random tensor networks. In these contexts, the entanglement entropy undergoes a transition between area-law and volume-law scaling, depending on the model parameters.
This transition point has attracted considerable attention.

In the present work, we investigate the ferromagnetic–paramagnetic phase transition of the permutation model, which corresponds to the entanglement entropy transition. 
Using exact real-space renormalization group calculations on self-dual hierarchical lattices, we numerically determine finite-replica critical points for \(q=mn=2,\ldots,6\). 
We compare the results with the duality prediction based on the Fourier transform of the symmetric group and then extrapolate the $b=3$ data toward the replica limit $mn\to0$, where the effective dimension is two. 
The comparison supports the duality-based estimate while also clarifying the systematic uncertainty associated with the extrapolation formula.}

\kword{permutation model, random tensor network, real-space renormalization group, hierarchical lattice, duality}

\begin{document}
\maketitle

\section{Introduction}
Current quantum devices are referred to as noisy intermediate-scale quantum (NISQ) computers, as they are subject to noise from environmental interactions, quantum gate operations, and measurement processes \cite{Preskill2018quantumcomputingin}. It is well established that when the noise level is high, quantum computers fail to demonstrate an advantage over classical computers \cite{PhysRevLett.69.2863_DMRG,PhysRevLett.101.250602_PEPS}. Conversely, in the low-noise regime, they can efficiently solve problems that are otherwise intractable for classical computation \cite{Arute2019,Lund_2017}. Thus, determining the noise threshold below which quantum computers retain their computational advantage is a crucial issue. 

Analyzing a random quantum circuit (RQC) provides insights into the noise tolerance of quantum computers. It has been shown that RQC can be mapped onto permutation models \cite{Fisher_2023,PhysRevB.101.104302_MIPT_rqc,PhysRevB.100.134306_MIPT_hybrid_qc}. In this mapping, the ordered and disordered phases of the effective classical model encode different entanglement structures of the underlying quantum dynamics. Identifying the ferromagnetic-to-paramagnetic phase transition point of the permutation model is thus expected to shed light on the noise threshold of NISQ devices.

The random tensor network (RTN) represents another class of quantum systems that can also be mapped onto permutation models. These models connect to broader fields such as quantum gravity and thermalization in quantum many-body systems \cite{Hayden_2016,Vasseur_2019}. In RTN, the ferromagnetic phase of the permutation model corresponds to one entanglement-scaling regime, while the paramagnetic phase corresponds to the other. The permutation models derived from RTN are generally simpler and more analytically tractable than those derived from RQC, making the study of phase transitions in RTN a useful step toward developing analytical approaches for RQC.

Duality analysis has a long history as a way of locating transition points in two-dimensional classical spin systems. In random spin systems, the combination of duality, gauge symmetry, and the replica method led to accurate conjectures for multicritical points on the Nishimori line. Subsequent studies showed both the strength and the limitation of this approach: exact RSRG on hierarchical lattices revealed small but meaningful deviations from the conventional single-equation conjecture \cite{Ohzeki_2008}, while improved formulations based on renormalized principal Boltzmann factors substantially reduced these discrepancies and were extended to regular lattices \cite{Ohzeki_2009_regular}. A further refinement combined duality analysis with real-space renormalization and graph-polynomial ideas, replacing fixed boundary conditions by periodic and twisted-periodic ones and achieving high-precision estimates of spin-glass phase diagrams \cite{Ohzeki_Jacobsen_2015}. These developments make it important to distinguish between the exact duality relation itself and the additional assumptions used to turn it into a scalar critical-point equation.

A previous study \cite{ohzeki2024duality} generalized this line of reasoning to permutation models by using the Fourier transform on the symmetric group. That study analyzed the phase transition point of RTN using duality analysis and predicted the critical point of the effective permutation model, followed by extrapolation to the RTN limit. However, because the reduction of the duality relation to a scalar critical-point condition involves nontrivial assumptions, independent validation by a different method is required. 

In this work, we analyze the phase transition point of the permutation model using the real-space renormalization group (RSRG) on hierarchical lattices \cite{PhysRevB.26.5022_RSRG_hier,Ohzeki_2008}. The real-space renormalization group is a powerful method for investigating phase transitions and critical phenomena in spin models. 
Hierarchical lattices are particularly useful because they allow for exact RSRG without approximation, thus providing a controlled numerical setting in which the flow of the Boltzmann weights can be followed directly. 
Self-duality still plays an essential role: if the scalar critical-point equation inferred from duality is exact, it should be compatible with exact RSRG results on self-dual hierarchical lattices. 
In other words, deviations quantify the limitation of reducing the full duality relation to a single equation. This is the sense in which the present RSRG calculation verifies the assumption made in prior work.

We compute the transition points for small values of a control parameter of our system and use extrapolation to estimate the critical point for RTN. 
Our results are generally consistent with the predicted RTN transition range in \cite{Vasseur_2019} and with the permutation-model prediction from \cite{ohzeki2024duality}. 
At the same time, comparing different hierarchical-lattice cells and extrapolation functions reveals the remaining systematic uncertainty, which is important because the RTN limit requires an analytic continuation of the parameter to zero.

The structure of this paper is as follows. 
Section \ref{sec:permutation_model} defines the permutation model under consideration and outlines the motivation for analyzing its phase transition points, focusing on the connection to RTN. We also review previous results based on duality analysis and highlight the challenges inherent in those studies. 
Section \ref{sec:renormalization} introduces hierarchical lattices and presents the theoretical framework for analyzing phase transitions using RSRG. 
Section \ref{sec:numerical_experiments} describes the computational procedure for applying RSRG to permutation models on hierarchical lattices and presents the numerical results obtained for three choices of the length parameter $b$. We then fit these results and extrapolate them to estimate the transition point of RTN. 
Finally, Appendix \ref{sec:fitting_function} discusses the choice of fitting functions used in extrapolating the RTN transition point. We compare polynomial, power-law, and rational-function forms and use the replicated random-bond Ising model on a hierarchical lattice as a benchmark for the reliability of replica extrapolation.

\section{Permutation model}\label{sec:permutation_model}

The permutation model is closely related to RTN, which are quantum states with inherent randomness defined on a network.  
Using the replica method to handle disorder in RTN, one finds that the entanglement entropy equals the difference in free energies of the \(q\)-th permutation model with specific boundary conditions, where the interaction strength is set by \(J\). Here, the order of the Rényi entropy is denoted by \(n\), while the number of replicas is \(m\), which is eventually taken to zero to extract the relevant information.  
In particular, the transition between volume-law and area-law scaling of entanglement entropy in RTN maps onto the ferromagnetic–paramagnetic phase transition in the permutation model. The detailed derivations of this correspondence can be found in \cite{Hayden_2016,Vasseur_2019}.

We write \(q=mn\) when it is useful to express the order of the symmetric group. 
Each spin \(\sigma_v\), located at a vertex \( v \in V \), is an element of the symmetric group \(\mathfrak{S}_{q}\). The energy of a spin configuration \(\bm{\sigma} = (\sigma_v)_{v \in V}\) is defined as
\begin{align}
H(\bm{\sigma}) &= -J \sum_{(i,j)\in E} C(\sigma_i\sigma_j^{-1}), \label{eq:hamiltonian_permutation_spin}
\end{align}
where \(E\) denotes the set of edges, \(J\) is the interaction strength, and \(C\) is the cycle-counting function on \(\mathfrak{S}_{q}\). The function \(C(\sigma)\) gives the number of disjoint cycles in the permutation \(\sigma\). For instance, the identity permutation \(e\) yields \(C(e) = q\), while for \(\sigma = (1,2)(3)\), we obtain \(C(\sigma) = 2\). By definition, \(C(\sigma)\) satisfies
\begin{align}
C(\sigma^{-1}) = C(\sigma).
\end{align}
Moreover, since the number of disjoint cycles is invariant under conjugation, \(C\) is a class function:
\begin{align}
C(\tau^{-1} \sigma \tau) = C(\sigma).
\end{align}
The interaction energy remains unchanged under global transformations of the spins \(\sigma \to \sigma\tau\) or \(\sigma \to \tau\sigma\), ensuring that the model possesses \( S_{q} \) symmetry.

Duality analysis is a widely used method to determine phase transition points in classical spin systems. 
Previous studies have applied this approach to predict the transition point of the \( q \)-th permutation model on a square lattice. By further taking the replica limit \( q \to 0 \), the corresponding transition point of the entanglement entropy in RTN was also predicted \cite{ohzeki2024duality}.

In the RTN normalization used below, the dimensionless coupling is expressed as \(K=\beta J=\log D\), where \(D\) is the bond dimension. 
We therefore quote the critical point as \(D_c=\exp(K_c)\). 
From duality analysis, the following equation is obtained:
\begin{align}
    D_c^{\,q} = \frac{1}{\sqrt{\Gamma(q+1)}} \, \frac{\Gamma(D_c+q)}{\Gamma(D_c)}. 
    \label{eq:critical_point_permutation_model_using_duality_analysis}
\end{align}
This equation determines the critical point \( D_c \) as a function of \( q \).  
Extrapolating to the replica limit \( q \to 0 \), we find
\begin{align}
    \log D_c \xrightarrow{q \to 0} \frac{\gamma}{2} + \psi(D_c),
\end{align}
where \(\psi(x) = \frac{\mathrm{d}}{\mathrm{d}x}\log \Gamma(x)\) is the digamma function, and \(\gamma = -\psi(1)\) is the Euler–Mascheroni constant. 
The resulting estimate is \( D_c \approx 1.882 \), which lies within the predicted range \( 1 < D_c < 2 \) reported in earlier studies \cite{Vasseur_2019}.  

This equation is obtained after reducing the full duality relation to a scalar condition for the principal Boltzmann factor.
Such a reduction is natural on self-dual lattices but is not automatic for a model with many independent Boltzmann weights. Independent validation of the result is therefore required.

\section{Real-space renormalization group on hierarchical lattice}\label{sec:renormalization}

To validate the predictions obtained from duality analysis, we analyze the phase transition points of the permutation model using the RSRG on hierarchical lattices. 
A key advantage of RSRG on hierarchical lattices is that it yields exact recursion relations for edge Boltzmann factors without generating uncontrolled long-range interactions \cite{PhysRevB.26.5022_RSRG_hier}. The comparison with duality is therefore especially transparent: the RSRG flow follows all effective edge weights generated by coarse graining. 
This strategy has also been successfully applied to spin-glass systems, where it clarified how accurate the conventional duality conjecture is and how it can be improved \cite{Ohzeki_2008}.

\begin{figure}[htbp]
\begin{center}
\includegraphics[width=\columnwidth]{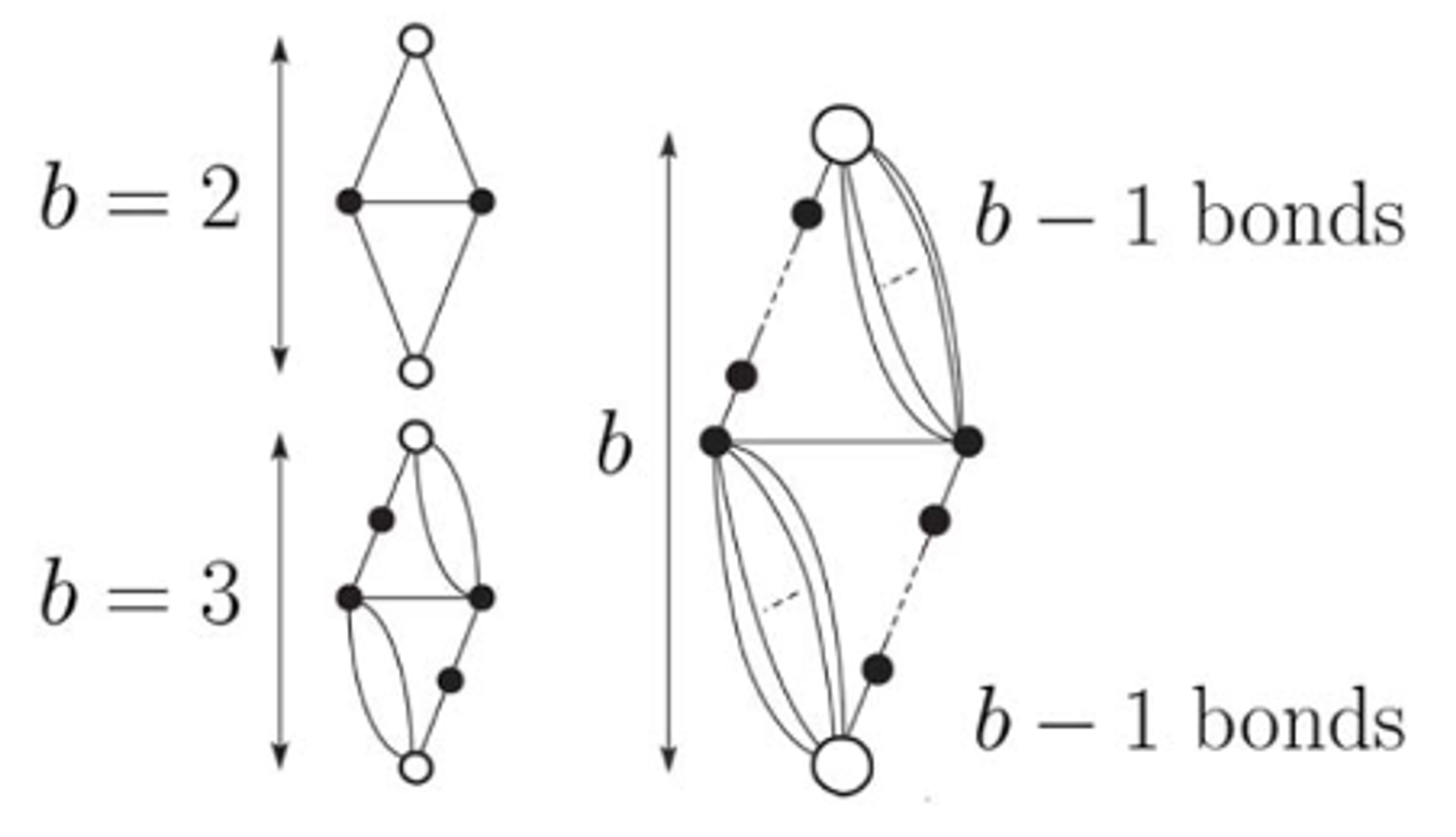}
\caption{Unit lattice. White circles denote boundary spins, and black circles denote internal spins to be summed over in the real-space renormalization.}
\label{fig:unit_lattice}
\end{center}
\end{figure}

\begin{figure}[htbp]
\begin{center}
\includegraphics[width=\columnwidth]{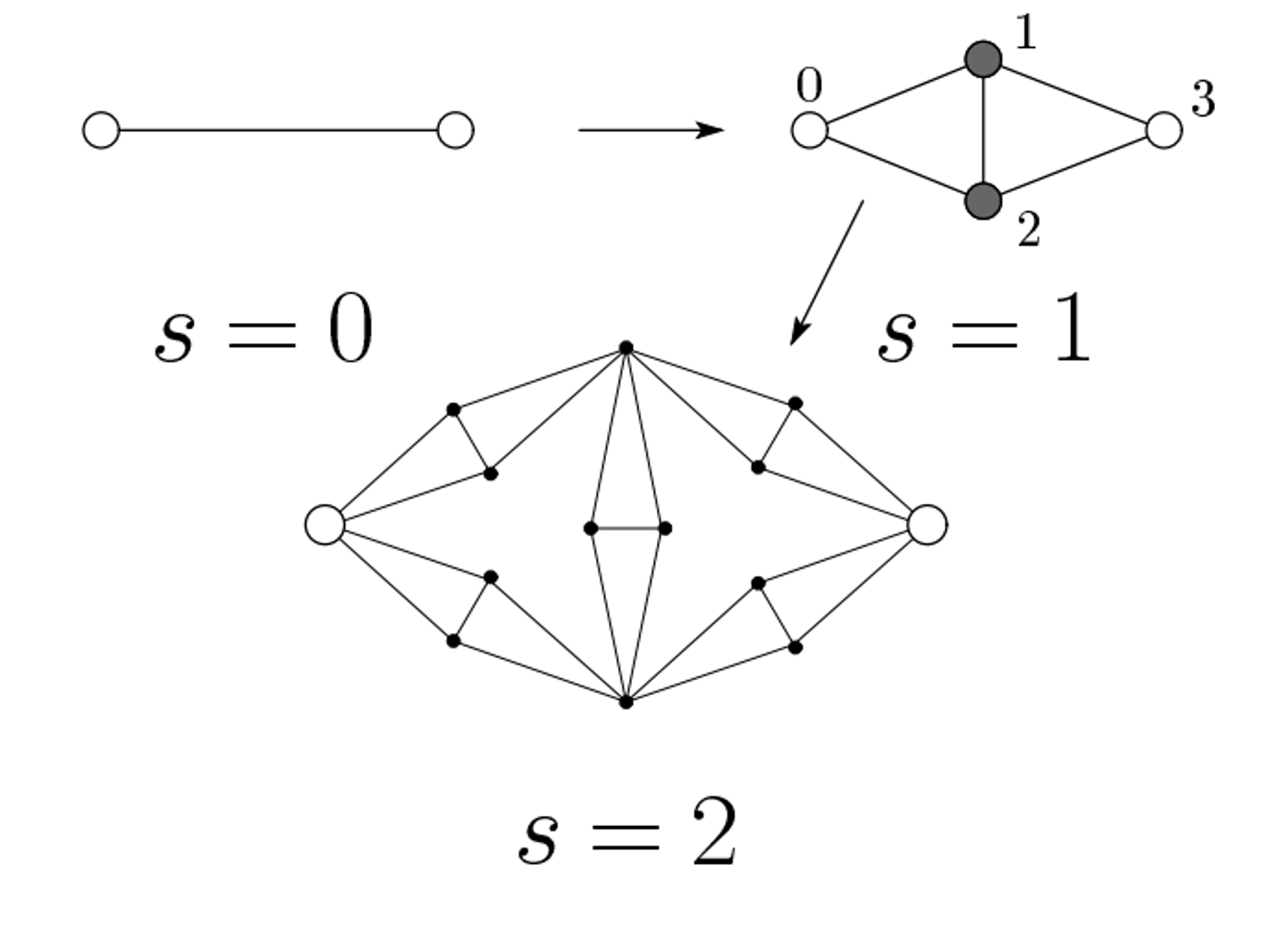}
\caption{Examples of hierarchical lattices with $b=2$. The iterative construction generates the entire lattice.}
\label{fig:ex_hierarchical_lattice}
\end{center}
\end{figure}

An \(r\)-level hierarchical lattice \(G_r = (V_r,E_r)\) is defined recursively by replacing each edge of an \((r-1)\)-level lattice with a unit cell. 
At \(r=0\), the lattice consists of two vertices connected by a single edge. Hierarchical lattices constructed from such unit cells are self-dual. 
In this study, we compute unit cells of length \(b=2\), \(b=3\), and \(b=4\). 
In particular, the case with \(b=3\) has a fractal dimension that coincides with two dimensions (\(d=2\)), making it quantitatively comparable to the square lattice as used in the literature \cite{Ohzeki_2008}. 
The $b=2$ and $4$ data are included as an additional check of the dependence on the hierarchical-lattice construction. 
The unit cells are shown in Fig.~\ref{fig:unit_lattice}, and examples of hierarchical lattices for \(r=0,1,2\) with \(b=2\) are shown in Fig.~\ref{fig:ex_hierarchical_lattice}.

In RSRG, spins on vertices in the \((r+1)\)-level lattice with inverse temperature \(\beta^{(r+1)}\) are summed out to obtain an effective model on the \(r\)-level lattice with inverse temperature \(\beta^{(r)}\). 
The partition function transforms as
\begin{align}
    Z_{r+1}(\beta^{(r+1)}) 
    &= \sum_{\bm{\sigma} \in \mathfrak{S}_{q}^{N_{r+1}}} e^{-\beta^{(r+1)} H_{r+1}(\bm{\sigma})} \\
    &= \sum_{\bm{\sigma} \in \mathfrak{S}_{q}^{N_r}}
    \left[ \sum_{\{\sigma_k\}_{k \in V_{r+1}\setminus V_r}} 
    e^{-\beta^{(r+1)} H_{r+1}(\bm{\sigma})} \right]_{\{\sigma_i\}_{i \in V_r}} \label{eq:pre_renoramlize}\\
    &= \sum_{\bm{\sigma} \in \mathfrak{S}_{q}^{N_r}} \Lambda^{(r)}(\beta^{(r)}) e^{-\beta^{(r)} H_r(\bm{\sigma})} \label{eq:post_renoramlize}\\
    &= \Lambda^{(r)}(\beta^{(r)}) Z_r(\beta^{(r)}). \label{eq:boltzmann_factor_origin}
\end{align}
The step from Eq.~\eqref{eq:pre_renoramlize} to Eq.~\eqref{eq:post_renoramlize} is the usual closed-coupling description, in which the functional form of the Boltzmann factor is assumed to be preserved under renormalization. 
In the numerical calculation below we keep the more general edge-weight vector generated by the right-hand side, so this closure assumption is used only for the simpler explanation of the spirit of RSRG.

The trivial fixed points of this transformation are \(\beta=0\) (paramagnetic) and \(\beta=\infty\) (ferromagnetic). 
A nontrivial fixed point, located between these two, corresponds to the phase transition. 
Thus, determining the critical point reduces to solving Eq.~\eqref{eq:boltzmann_factor_origin} under the fixed-point condition \(\beta^{(r)}=\beta^{(r+1)}\).

In the present calculation, we do not have a closed-form expression for the Boltzmann factor at each step of RSRG.
After one renormalization step, the effective edge weight is treated as a general function of the relative permutation. 
The initial condition is the one-parameter family \(e^{KC(\sigma)}\), but the effective edge weight \(\vec{\rho}\) itself is the running object under RSRG.
The update equation is
\begin{align}
    \Lambda_{r+1}(K)\, \rho_{r+1}(\sigma_i,\sigma_j|K) = 
    \left[ \sum_{\sigma_1,\sigma_2} \prod_{(k,l)\in E_{\mathrm{unit}}} 
    \rho_r(\sigma_k,\sigma_l|K) \right]_{\sigma_i,\sigma_j}. \label{eq:RSRG_permutation_model}
\end{align}
where $\rho_r(\sigma_k,\sigma_l|K)$ is the relative edge weight at each step $r$.
In addition, $\Lambda_{r}(K)$ is a normalization factor associated with the identity sector \(\sigma_i\sigma_j^{-1}=e\).
Successive renormalizations yield updated Boltzmann factors, which eventually flow to trivial fixed points corresponding either to the ferromagnetic or paramagnetic phases. 
We define the ferromagnetic phase by $\rho_r(\cdot|K) = (1,0,0,\ldots)$, where the first component corresponds to $\sigma_i\sigma_j^{-1} = e$, and the paramagnetic phase by $\rho_r(\cdot|K) = (1,1,1,\ldots)/q!$.
The critical point \(K_c\) is identified by tracking the separatrix between flows from the initial coupling toward distinct phases. 
To perform the iteration, we represent the Boltzmann factor as a vector indexed by \(\sigma \in \mathfrak{S}_{q}\):
\begin{align}
    \vec{\rho}(K) = \bigl(\rho(\sigma|K)\bigr)_{\sigma \in \mathfrak{S}_{q}}.
\end{align}
The initial condition is given as
\begin{equation}
    \rho(\sigma|K) = e^{K C(\sigma)}.
\end{equation}
Gauge invariance under global transformations \(\sigma \to \tau\sigma\) ensures that the pair weight depends only on the relative permutation \(\sigma_i\sigma_j^{-1}\). 
Thus the vector has size \(|\mathfrak{S}_{q}| = q!\), rather than \(|\mathfrak{S}_{q}|^2\). 
Since \(C(\sigma)\) is a class function and the unit-cell summation preserves conjugacy invariance, the weights may also be grouped by conjugacy classes. 
This class representation is used for \(q=5,6\), where storing all permutations explicitly becomes inefficient.

The direction of the RG flow is diagnosed by the weight of the identity permutation. 
The ferromagnetic fixed point concentrates the normalized weight at \(\sigma=e\), whereas the paramagnetic fixed point approaches the uniform distribution. 
For a given bond dimension \(D=e^K\), we therefore compute
\begin{align}
    G_N(D) = \rho_N (e|D)-\rho_0 (e|D),
\end{align}
after a sufficient number of RG steps.
In our study, we set \(N=50\). 
A positive value of \(G_N(D)\) indicates that the flow enhances the ferromagnetic component, while a negative value indicates flow toward the paramagnetic side.

The critical bond dimension \(D_c^{(q)}\) is determined using a binary search. 
We initialize an interval with \(D_l=1\) and \(D_u=10\), enlarge \(D_u\) if necessary until \(G_N(D_u)>0\), and set \(D=(D_l+D_u)/2\). 
If \(G_N(D)>0\), the upper bound is updated \(D_u \gets D\); otherwise the lower bound is updated \(D_l \gets D\). 
The process is repeated until the interval width satisfies \(D_u-D_l < 10^{-12}\), at which point the critical point is estimated as the midpoint:
\begin{align}
    D_c = \frac{D_l + D_u}{2}, \qquad K_c=\log D_c.
\end{align}

\section{Numerical experiments}\label{sec:numerical_experiments}
\begin{figure*}[t]
\begin{center}
\includegraphics[width=0.72\textwidth]{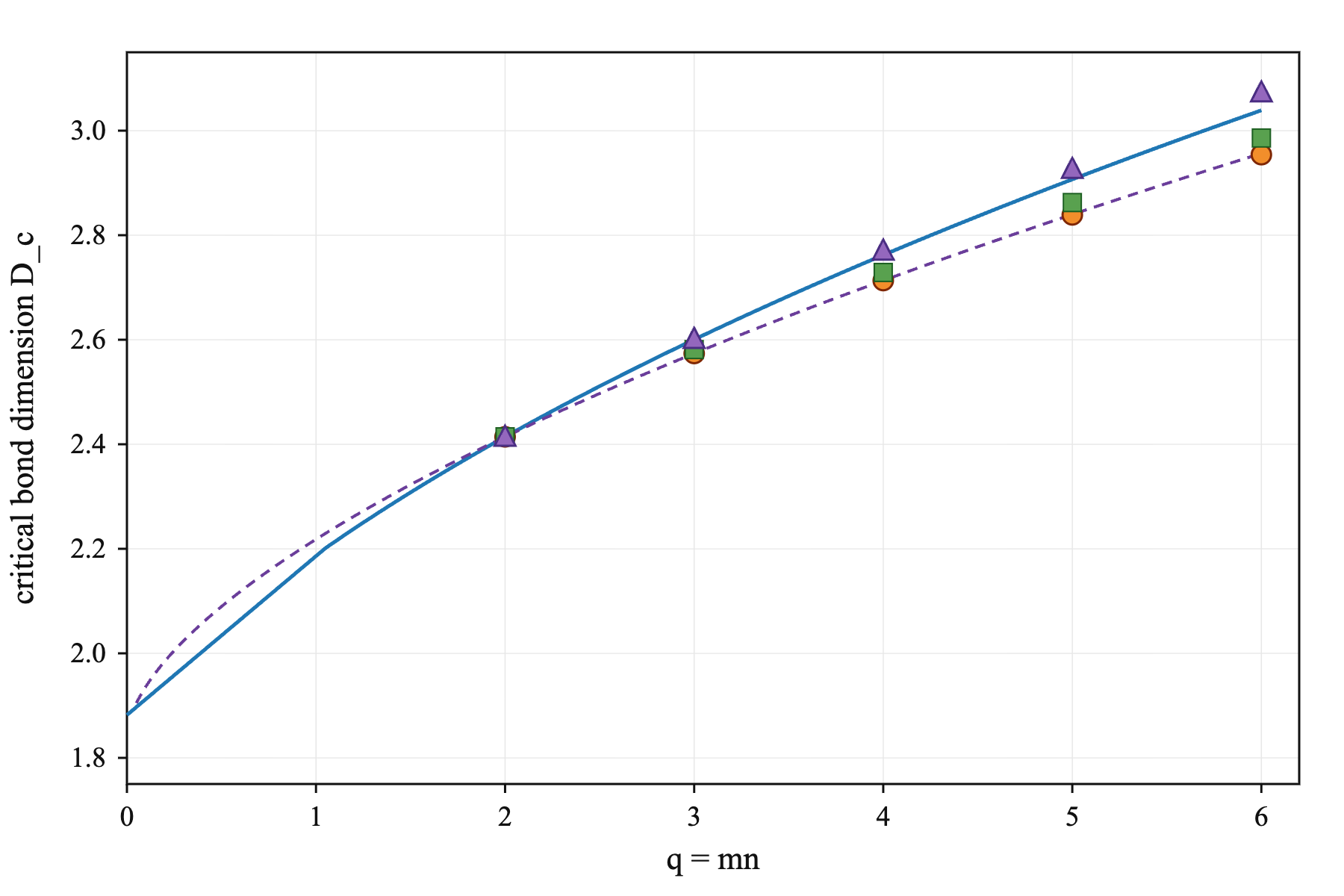}
\caption{(Color online) Critical points obtained by RSRG on hierarchical lattices. Orange circles, green squares, and purple triangles denote the RSRG results for $b=2,3,4$, respectively. The blue solid curve denotes the duality-analysis prediction, and the purple dashed curve is a power-law guide to the $b=2$ data. The replica-limit estimates discussed in the text are obtained from the $b=3$ data.}
\label{fig:critical_point_permutation}
\end{center}
\end{figure*}

\begin{figure*}[t]
\begin{center}
\includegraphics[width=0.72\textwidth]{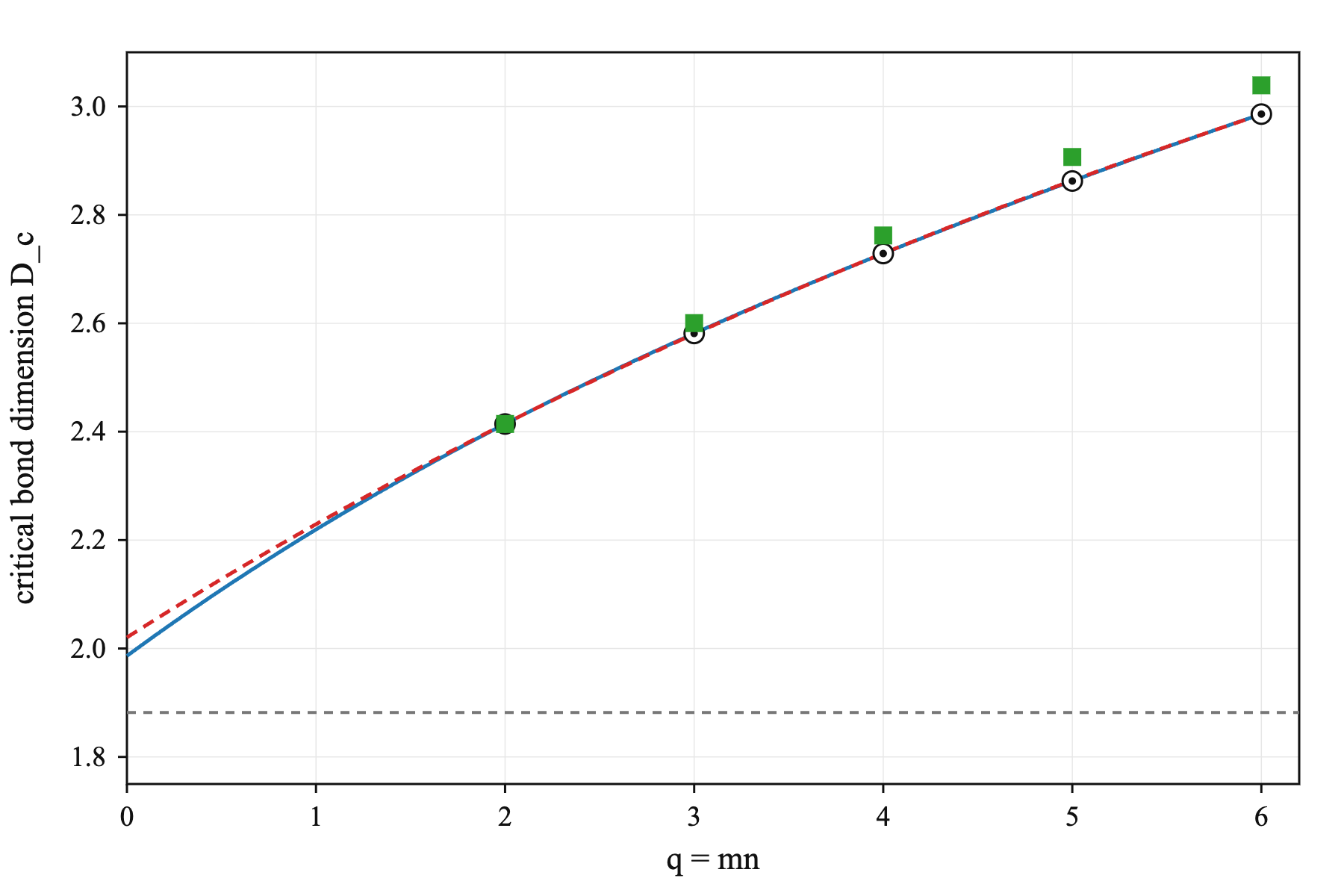}
\caption{(Color online) Extrapolation of the $b=3$ RSRG critical points. 
Open circles denote RSRG results, green squares denote the finite-\(q\) duality-analysis values, the blue solid curve is the fourth-order polynomial fit, the red dashed curve is the rational-function fit, and the gray dashed horizontal line is Ohzeki's duality estimate in the replica limit, $D_c^{(\mathrm{RTN})}=1.88201$.}
\label{fig:rsrg_b3_extrapolation}
\end{center}
\end{figure*}

The critical points of the $q$-th permutation models obtained using RSRG analysis are plotted in Figures \ref{fig:critical_point_permutation} and \ref{fig:rsrg_b3_extrapolation}, and shown in Table \ref{tab:critical_point_permutation}. 
In Figure \ref{fig:critical_point_permutation}, the curve represented by ``duality analysis'' corresponds to the predicted critical points, which were numerically solved from Eq.~\ref{eq:critical_point_permutation_model_using_duality_analysis} for real $q$. 
The point at $q=1$ is not plotted because Eq.~\ref{eq:critical_point_permutation_model_using_duality_analysis} becomes an identity and does not determine a unique transition point. 
The dashed curve for the $b=2$ data is included only as a visual guide; the replica extrapolation used below is based on the $b=3$ data and is compared among several functional forms.

\begin{table*}[t]
    \centering
    \begin{tabular}{c|c|c|c|c}
         $q$ & \multicolumn{4}{c}{$D_c$} \\ \hline
         {} & \shortstack{Hierarchical lattice\\duality analysis}
           & \shortstack{Hierarchical lattice($b=2$)\\RSRG}
           & \shortstack{Hierarchical lattice($b=3$)\\RSRG}
           & \shortstack{Hierarchical lattice($b=4$)\\RSRG}\\ \hline\hline
         2 & 2.41421 & 2.41421 & 2.41421 & 2.41421\\ 
         3 & 2.60032 & 2.57369 & 2.58121 & 2.60133\\ 
         4 & 2.76203 & 2.71325 & 2.72869 & 2.77043\\
         5 & 2.90679 & 2.83887 & 2.86247 & 2.92671\\
         6 & 3.03893 & 2.95408 & 2.98593 & 3.07304
    \end{tabular}
    \caption{Critical points of the \(q\)-order permutation spin models.}
    \label{tab:critical_point_permutation}
\end{table*}

The additional $b=4$ data show a stronger dependence on the hierarchical-lattice construction than the $b=2$ and $b=3$ results. 
In particular, the $b=4$ values lie above the $b=3$ values for $q\ge 3$ and are slightly larger than the finite-$q$ duality estimates for $q\ge 3$. 
Since the $b=3$ lattice has the two-dimensional fractal dimension relevant for comparison with the square lattice, the replica extrapolation below is based on the $b=3$ data, while the $b=4$ data are used as a check of lattice-construction dependence.

The critical point for the RTN is obtained by extrapolating the finite-$q$ data to the replica limit \(q\to0\). 
This is the most delicate step of the analysis, because the RSRG calculation is performed at positive integer values of \(q\), whereas the quenched RTN limit requires analytic continuation to zero replicas. 
For the $b=3$ data, we examined several extrapolation functions, following the same strategy used to test replica extrapolations in the random-bond Ising model: polynomial fits of different degrees, a power function, and a rational function. 
The rational function was defined as
\begin{align}
    F_{\mathrm{rat}}(q) = \frac{c + a q}{1 + \lambda q},
\end{align}
so that the desired replica-limit estimate is the intercept \(F_{\mathrm{rat}}(0)=c\). 
The comparison is shown in Table \ref{tab:rsrg_b3_extrapolation}, where the reference value is the duality estimate $D_c^{(\mathrm{RTN})}=1.88201$.
The table also lists the root-mean-square fitting error on the finite-\(q\) data,
\begin{align}
    \epsilon_{\mathrm{fit}} =
    \sqrt{\frac{1}{N_{\mathrm{data}}}\sum_i\left(F(q_i)-D_c^{(q_i)}\right)^2},
\end{align}
with \(q_i=2,\ldots,6\).

\begin{table*}[t]
    \centering
    \begin{tabular}{c|c|c|c|c}
        Fitting function & $D_c^{(\mathrm{RTN})}$ & \shortstack{Difference from\\duality analysis} & Relative difference & \(\epsilon_{\mathrm{fit}}\) \\ \hline\hline
        Linear polynomial & 2.14463 & 0.26262 & 13.95\% & $1.21\times10^{-2}$\\
        Quadratic polynomial & 2.04384 & 0.16183 & 8.60\% & $1.31\times10^{-3}$\\
        Cubic polynomial & 2.00521 & 0.12320 & 6.55\% & $1.29\times10^{-4}$\\
        Fourth-order polynomial & 1.98631 & 0.10430 & 5.54\% & $2.67\times10^{-13}$\\
        Power function & 1.82993 & $-0.05208$ & 2.77\% & $1.85\times10^{-4}$\\
        Rational function & 2.02040 & 0.13839 & 7.35\% & $6.89\times10^{-4}$
    \end{tabular}
    \caption{Comparison of \(q \to 0\) extrapolations for the $b=3$ RSRG data at \(q=2,\ldots,6\). The fitting error \(\epsilon_{\mathrm{fit}}\) is the RMSE on these five finite-\(q\) data points.}
    \label{tab:rsrg_b3_extrapolation}
\end{table*}

Among the polynomial fits, increasing the degree systematically reduces the finite-\(q\) fitting error, as expected from the interpolation argument. 
In particular, the fourth-order polynomial almost interpolates the five data points and therefore has an extremely small \(\epsilon_{\mathrm{fit}}\). 
This does not by itself guarantee the most reliable \(q\to0\) extrapolation. 
The fourth-order polynomial gives $D_c^{(\mathrm{RTN})}=1.98631$, which is above Ohzeki's duality estimate by $5.54\%$. 
The rational function gives $D_c^{(\mathrm{RTN})}=2.02040$, corresponding to a $7.35\%$ relative difference. 
These two fits therefore give values in the same qualitative range, but both overestimate the duality result for the present RSRG data. 
The power-function extrapolation gives the smallest difference in this data set, but its stability is less directly motivated than the polynomial expansion around \(q=0\). 
The fourth-order estimate satisfies the expected range $1 < D_c^{(\mathrm{RTN})} < 2$ \cite{Vasseur_2019}, while the rational-function estimate lies slightly above its upper edge. 
We therefore avoid selecting a single value based solely on the smallest fitting error and instead use the spread among these fits as an estimate of the present extrapolation uncertainty.

To assess whether the extrapolation functions are reasonable, we compared them with the finite-replica benchmark available for the random-bond Ising model on self-dual hierarchical lattices \cite{Ohzeki_2008}. 
In that problem, the $n\to0$ limit can be obtained directly, and finite-replica results for $n=1,\ldots,4$ are also available. 
As summarized in Appendix \ref{sec:fitting_function}, the $b=3$ benchmark strongly favors the rational form over low-degree polynomial fits when one extrapolates from $n=1,\ldots,4$ to $n=0$. 
We also tested the power-function form in the same benchmark. 
Although it fits the finite-$n$ points reasonably well, the unconstrained optimal exponent is negative, so the function has no finite $n\to0$ limit. 
This indicates that the apparent success of the power function in the present permutation-model data should be interpreted cautiously. 
A fourth-order polynomial cannot be independently tested from only four finite-replica points in the benchmark. 
We therefore regard the rational fit as empirically motivated by the spin-glass replica analysis, whereas the power function and the fourth-order polynomial should be interpreted as complementary empirical checks rather than universally justified extrapolation formulas.

\section{Conclusion}\label{sec:conclusion}
In this paper, we numerically determined the ferromagnetic-paramagnetic phase transition points of permutation spin models related to random quantum circuits and random tensor networks. Using the real-space renormalization group (RSRG) method on hierarchical lattices, we determined the critical points for \(q=2,\ldots,6\) and, by extrapolating these to \(q \to 0\), estimated the volume-law to area-law phase transition point for the entanglement entropy of random tensor networks.

The finite-\(q\) comparison shows that the exact RSRG results are broadly consistent with the duality prediction in Eq.~\ref{eq:critical_point_permutation_model_using_duality_analysis}, especially for the $b=2$ and $b=3$ hierarchical lattices. This supports the usefulness of the duality-based estimate for the permutation model. At the same time, the $b=4$ data and the replica extrapolation show that the agreement should be interpreted with care: the full RG flow involves many effective Boltzmann weights, whereas the duality argument reduces the problem to a scalar condition.

The comparison of several extrapolation functions also clarifies the present limitation of the numerical analysis. The rational function is supported by the benchmark from the replicated random-bond Ising model on hierarchical lattices, where direct $n\to0$ results are known. The power function gives the closest value to the duality estimate in the present $b=3$ permutation-model data, but the same form is unstable in the random-bond Ising benchmark because the optimized exponent becomes negative. The fourth-order polynomial gives a value within the expected RTN interval, but it should be regarded as a smooth interpolation of the finite-$q$ data rather than an independently established replica-extrapolation scheme. Additional finite-$q$ data, or a formulation that accesses the replica limit more directly, would be needed to decide more sharply which extrapolation form is most reliable for the permutation model.

Looking ahead, one potential direction is the development of methods for determining the phase transition points of permutation spin models corresponding to random quantum circuits. 
The effective spin model corresponding to random quantum circuits is a permutation spin model defined on a hexagonal lattice. 
By utilizing duality analysis with star-triangle transformations, we expect to predict the phase transition points analytically. 
While these predictions should be numerically verified, it should be noted that since the hexagonal lattice is not a self-dual graph, it cannot be verified using real-space renormalization group methods on hierarchical lattices. 
Therefore, it is essential to apply renormalization group techniques suitable for models on hexagonal lattices.

\begin{acknowledgment}
We received financial support from the Cross-ministerial Strategic Innovation Promotion Program (SIP) of the Cabinet Office (No. 23836436).
\end{acknowledgment}

\appendix

\section{Fitting functions}\label{sec:fitting_function}
In this section, we summarize the fitting functions used to determine the phase transition point of RTN from finite-\(q\) data. Let \(q=mn\) and denote the RSRG critical bond dimension obtained at finite \(q\) by \(D_c^{(q)}\). We fit these data by a function \(F(q)\), and the replica-limit estimate is obtained from its intercept:
\begin{align}
    D_c^{(\mathrm{RTN})} \simeq F(0).
\end{align}

The polynomial fits are written as
\begin{align}
    F_{\mathrm{poly}}^{(p)}(q) = a_0 + a_1 q + a_2 q^2 + \cdots + a_p q^p,
\end{align}
where \(p=1,2,3,4\) in the numerical comparison. This form corresponds to a truncated Taylor expansion around \(q=0\), and the estimate is \(F_{\mathrm{poly}}^{(p)}(0)=a_0\).

We also compare the power-function fit
\begin{align}
    F_{\mathrm{pow}}(q) = a q^\alpha + c,
\end{align}
for which \(F_{\mathrm{pow}}(0)=c\) if \(\alpha>0\), and the rational-function fit
\begin{align}
    F_{\mathrm{rat}}(q) = \frac{c + a q}{1 + \lambda q},
\end{align}
for which the intercept is again \(F_{\mathrm{rat}}(0)=c\). For the power function, the identification of \(c\) with the replica-limit value is meaningful only when the fitted exponent satisfies \(\alpha>0\). We use \(\alpha\) and \(\lambda\) here to avoid confusion with the hierarchical-lattice length parameter \(b\). The rational form is included because it was found to be a useful empirical extrapolation function in the replica analysis of the random-bond Ising model.

We checked this point using the benchmark data of the replicated random-bond Ising model on self-dual hierarchical lattices \cite{Ohzeki_2008}. In that model, the quenched limit $n\to0$ can be evaluated directly, while exact finite-replica RSRG data are available for $n=1,\ldots,4$. This makes it possible to test how well an extrapolation from finite $n$ reproduces the known $n\to0$ result. 
For the $b=3$ self-dual hierarchical lattice, the direct estimate is $p_{\rm numerical}(n\to0)=0.8903(2)$. Fitting the finite-replica values at $n=1,\ldots,4$ gives the results in Table \ref{tab:rbim_replica_benchmark}.

\begin{table*}[t]
    \centering
    \begin{tabular}{c|c|c|c}
        Fitting function & $p_{\rm ext}(n\to0)$ & $p_{\rm ext}-0.8903$ & \shortstack{RMSE on\\$n=1,\ldots,4$} \\ \hline\hline
        Linear polynomial & 0.83816 & $-0.05214$ & $5.07\times10^{-3}$\\
        Quadratic polynomial & 0.86322 & $-0.02708$ & $7.71\times10^{-4}$\\
        Cubic polynomial & 0.87528 & $-0.01502$ & $<10^{-12}$\\
        Power function & \multicolumn{2}{c|}{not finite at $n\to0$ ($\alpha=-0.0286$)} & $1.78\times10^{-4}$\\
        Rational function & 0.88915 & $-0.00115$ & $1.29\times10^{-5}$\\
        Fourth-order polynomial & \multicolumn{3}{c}{not determined from four finite-$n$ points}
    \end{tabular}
    \caption{Benchmark extrapolation for the replicated random-bond Ising model on the $b=3$ self-dual hierarchical lattice, using the finite-replica numerical values at $n=1,\ldots,4$ reported in Ref.~\cite{Ohzeki_2008}.}
    \label{tab:rbim_replica_benchmark}
\end{table*}

This benchmark has three implications for the present RTN extrapolation. First, the rational function can capture the curvature toward the replica limit much more efficiently than low-degree polynomials in a closely related hierarchical-lattice replica problem. 
Second, the power function is not robust in this benchmark: the unconstrained least-squares optimum gives \(\alpha=-0.0286\), so \(a n^\alpha+c\) diverges as \(n\to0\), and the fitted value \(c=-0.8773\) cannot be interpreted as a probability or as a replica-limit estimate. If one enforces \(\alpha>0\), the fit is driven toward the boundary \(\alpha\to0\) and the intercept becomes unstable. 
Thus, the small finite-\(n\) RMSE of the power function does not by itself justify its use for replica extrapolation. 
Third, a fourth-order polynomial cannot be validated by the finite-replica data of Ref.~\cite{Ohzeki_2008}, because only four nonzero replica numbers are available there. 
In the present permutation-model analysis, we have five finite data points, \(q=2,\ldots,6\), so the fourth-order polynomial is mathematically well defined, but it should be regarded as an interpolation assumption. The benchmark does not prove that the same rational form must be optimal for the permutation model; it only provides evidence that a rational extrapolation is a reasonable candidate in replica problems on hierarchical lattices. 
For this reason, we report the rational-function, power-function, and fourth-order polynomial estimates as complementary checks rather than selecting one of them solely by the finite-\(q\) fitting error.

\bibliographystyle{jpsj}
\bibliography{renormalization_permutation}

@article{ohzeki2024duality,
  title={Duality analysis in a symmetric group and its application to random tensor network models},
  author={Ohzeki, Masayuki},
  journal={Progress of Theoretical and Experimental Physics},
  doi = {10.1093/ptep/ptae171},
url = {https://doi.org/10.1093/ptep/ptae171},
  pages={ptae171},
  year={2024},
  publisher={Oxford University Press}
}

@article{Vasseur_2019,
   title={Entanglement transitions from holographic random tensor networks},
   volume={100},
   ISSN={2469-9969},
   url={http://dx.doi.org/10.1103/PhysRevB.100.134203},
   DOI={10.1103/physrevb.100.134203},
   number={13},
   journal={Physical Review B},
   publisher={American Physical Society (APS)},
   author={Vasseur, Romain and Potter, Andrew C. and You, Yi-Zhuang and Ludwig, Andreas W. W.},
   year={2019},
   month=oct }

@article{Preskill2018quantumcomputingin,
  doi = {10.22331/q-2018-08-06-79},
  url = {https://doi.org/10.22331/q-2018-08-06-79},
  title = {Quantum {C}omputing in the {NISQ} era and beyond},
  author = {Preskill, John},
  journal = {{Quantum}},
  issn = {2521-327X},
  publisher = {{Verein zur F{\"{o}}rderung des Open Access Publizierens in den Quantenwissenschaften}},
  volume = {2},
  pages = {79},
  month = aug,
  year = {2018}
}

@article{Lund_2017,
   title={Quantum sampling problems, BosonSampling and quantum supremacy},
   volume={3},
   ISSN={2056-6387},
   url={http://dx.doi.org/10.1038/s41534-017-0018-2},
   DOI={10.1038/s41534-017-0018-2},
   number={1},
   journal={npj Quantum Information},
   publisher={Springer Science and Business Media LLC},
   author={Lund, A. P. and Bremner, Michael J. and Ralph, T. C.},
   year={2017},
   month=apr }

@article{Arute2019,
  author = {Arute, Frank and Arya, Kunal and Babbush, Ryan and Bacon, Dave and Bardin, Joseph C. and Barends, Rami and Biswas, Rupak and Boixo, Sergio and Brandao, Fernando G. S. L. and Buell, David A. and Burkett, Brian and Chen, Yu and Chen, Zijun and Chiaro, Ben and Collins, Roberto and Courtney, William and Dunsworth, Andrew and Farhi, Edward and Foxen, Brooks and Fowler, Austin and Gidney, Craig and Giustina, Marissa and Graff, Rob and Guerin, Keith and Habegger, Steve and Harrigan, Matthew P. and Hartmann, Michael J. and Ho, Alan and Hoffmann, Markus and Huang, Trent and Humble, Travis S. and Isakov, Sergei V. and Jeffrey, Evan and Jiang, Zhang and Kafri, Dvir and Kechedzhi, Kostyantyn and Kelly, Julian and Klimov, Paul V. and Knysh, Sergey and Korotkov, Alexander and Kostritsa, Fedor and Landhuis, David and Lindmark, Mike and Lucero, Erik and Lyakh, Dmitry and Mandrà, Salvatore and McClean, Jarrod R. and McEwen, Matthew and Megrant, Anthony and Mi, Xiao and Michielsen, Kristel and Mohseni, Masoud and Mutus, Josh and Naaman, Ofer and Neeley, Matthew and Neill, Charles and Niu, Murphy Yuezhen and Ostby, Eric and Petukhov, Andre and Platt, John C. and Quintana, Chris and Rieffel, Eleanor G. and Roushan, Pedram and Rubin, Nicholas C. and Sank, Daniel and Satzinger, Kevin J. and Smelyanskiy, Vadim and Sung, Kevin J. and Trevithick, Matthew D. and Vainsencher, Amit and Villalonga, Benjamin and White, Theodore and Yao, Z. Jamie and Yeh, Ping and Zalcman, Adam and Neven, Hartmut and Martinis, John M.},
  title = {Quantum supremacy using a programmable superconducting processor},
  journal = {Nature},
  year = {2019},
  volume = {574},
  number = {7779},
  pages = {505--510},
  doi = {10.1038/s41586-019-1666-5},
  url = {https://doi.org/10.1038/s41586-019-1666-5}
}

@article{PhysRevLett.69.2863_DMRG,
  title = {Density matrix formulation for quantum renormalization groups},
  author = {White, Steven R.},
  journal = {Phys. Rev. Lett.},
  volume = {69},
  issue = {19},
  pages = {2863--2866},
  numpages = {0},
  year = {1992},
  month = {Nov},
  publisher = {American Physical Society},
  doi = {10.1103/PhysRevLett.69.2863},
  url = {https://link.aps.org/doi/10.1103/PhysRevLett.69.2863}
}

@article{PhysRevLett.101.250602_PEPS,
  title = {Classical Simulation of Infinite-Size Quantum Lattice Systems in Two Spatial Dimensions},
  author = {Jordan, J. and Or\'us, R. and Vidal, G. and Verstraete, F. and Cirac, J. I.},
  journal = {Phys. Rev. Lett.},
  volume = {101},
  issue = {25},
  pages = {250602},
  numpages = {4},
  year = {2008},
  month = {Dec},
  publisher = {American Physical Society},
  doi = {10.1103/PhysRevLett.101.250602},
  url = {https://link.aps.org/doi/10.1103/PhysRevLett.101.250602}
}

@article{Fisher_2023,
   title={Random Quantum Circuits},
   volume={14},
   ISSN={1947-5462},
   url={http://dx.doi.org/10.1146/annurev-conmatphys-031720-030658},
   DOI={10.1146/annurev-conmatphys-031720-030658},
   number={1},
   journal={Annual Review of Condensed Matter Physics},
   publisher={Annual Reviews},
   author={Fisher, Matthew P.A. and Khemani, Vedika and Nahum, Adam and Vijay, Sagar},
   year={2023},
   month=mar, pages={335–379} }

@article{PhysRevB.101.104302_MIPT_rqc,
  title = {Measurement-induced criticality in random quantum circuits},
  author = {Jian, Chao-Ming and You, Yi-Zhuang and Vasseur, Romain and Ludwig, Andreas W. W.},
  journal = {Phys. Rev. B},
  volume = {101},
  issue = {10},
  pages = {104302},
  numpages = {11},
  year = {2020},
  month = {Mar},
  publisher = {American Physical Society},
  doi = {10.1103/PhysRevB.101.104302},
  url = {https://link.aps.org/doi/10.1103/PhysRevB.101.104302}
}

@article{PhysRevB.100.134306_MIPT_hybrid_qc,
  title = {Measurement-driven entanglement transition in hybrid quantum circuits},
  author = {Li, Yaodong and Chen, Xiao and Fisher, Matthew P. A.},
  journal = {Phys. Rev. B},
  volume = {100},
  issue = {13},
  pages = {134306},
  numpages = {26},
  year = {2019},
  month = {Oct},
  publisher = {American Physical Society},
  doi = {10.1103/PhysRevB.100.134306},
  url = {https://link.aps.org/doi/10.1103/PhysRevB.100.134306}
}

@article{Hayden_2016,
   title={Holographic duality from random tensor networks},
   volume={2016},
   ISSN={1029-8479},
   url={http://dx.doi.org/10.1007/JHEP11(2016)009},
   DOI={10.1007/jhep11(2016)009},
   number={11},
   journal={Journal of High Energy Physics},
   publisher={Springer Science and Business Media LLC},
   author={Hayden, Patrick and Nezami, Sepehr and Qi, Xiao-Liang and Thomas, Nathaniel and Walter, Michael and Yang, Zhao},
   year={2016},
   month=nov }

@article{PhysRevB.26.5022_RSRG_hier,
  title = {Spin systems on hierarchical lattices. Introduction and thermodynamic limit},
  author = {Griffiths, Robert B. and Kaufman, Miron},
  journal = {Phys. Rev. B},
  volume = {26},
  issue = {9},
  pages = {5022--5032},
  numpages = {0},
  year = {1982},
  month = {Nov},
  publisher = {American Physical Society},
  doi = {10.1103/PhysRevB.26.5022},
  url = {https://link.aps.org/doi/10.1103/PhysRevB.26.5022}
}

@article{Ohzeki_2008,
   title={Multicritical points for spin-glass models on hierarchical lattices},
   volume={77},
   ISSN={1550-2376},
   url={http://dx.doi.org/10.1103/PhysRevE.77.061116},
   DOI={10.1103/physreve.77.061116},
   number={6},
   journal={Physical Review E},
   publisher={American Physical Society (APS)},
   author={Ohzeki, Masayuki and Nishimori, Hidetoshi and Berker, A. Nihat},
   year={2008},
   month=jun }

@article{Ohzeki_2009_regular,
  title={Locations of multicritical points for spin glasses on regular lattices},
  author={Ohzeki, Masayuki},
  journal={Physical Review E},
  volume={79},
  pages={021129},
  year={2009},
  doi={10.1103/PhysRevE.79.021129},
  eprint={0811.0464},
  archivePrefix={arXiv}
}

@article{Ohzeki_Jacobsen_2015,
  title={High-precision phase diagram of spin glasses from duality analysis with real-space renormalization and graph polynomials},
  author={Ohzeki, Masayuki and Jacobsen, Jesper Lykke},
  journal={Journal of Physics A: Mathematical and Theoretical},
  volume={48},
  number={9},
  pages={095001},
  year={2015},
  doi={10.1088/1751-8113/48/9/095001},
  eprint={1410.0166},
  archivePrefix={arXiv}
}

\end{document}